\documentclass[%
 reprint,
 amsmath,amssymb,
 aps,
]{revtex4-1}

\usepackage{physics}
\usepackage{amsmath, amsthm, amssymb,commath, calrsfs, wasysym, dsfont ,verbatim, bbm, color,  geometry,graphicx}
\usepackage{xcolor}
\usepackage{graphicx}
\usepackage{dcolumn}
\usepackage{bm}

\newcommand\authormark[1]{\textsuperscript{#1}}

\begin{document}

\preprint{APS/123-QED}

\title{Frequency modulation of Rydberg states by radio frequency electromagnetic fields}

\author{Nabendu S. Mishra\authormark{1,2,}}
\email{nmishra2@ur.rochester.edu}
\address{\authormark{1}{The Institute of Optics, University of Rochester, 480 Intercampus Drive, River Campus, Rochester, NY 14627}\\
\authormark{2}{Department of Physical Sciences, Indian Institute of Science Education and Research Berhampur, Ganjam, Odisha 760010, India}}

\vspace{3em}
\date{\today}

\begin{abstract}
 In this work, we demonstrate the generation of sidebands in electromagnetically induced transparency (EIT) signal due to frequency modulation of Rydberg states in $^{87}\text{Rb}$ thermal vapor. An oscillating radio-frequency (RF) electric field generated through a pair of parallel-placed copper plates leads to modulation of the Rydberg state by virtue of Stark effect where the frequency and strength of the sidebands quantify the RF electric field present. The variation of the strength of the EIT peak and the sidebands with RF voltage is observed to vary as a function of the Bessel function of modulation index. It provides an estimate of the atomic polarizability in the medium, which for $54S_{1/2}$ state, is found to be $(h)\cdot1.01\times10^8$ $\text{Hz/(V/cm)}^2$.

\end{abstract}

\maketitle

\section{Introduction}\label{sec1}
In recent times, the frontier of quantum science and technologies has emerged as one of the most trending areas in the world of scientific affairs. The latest developments and advancements in quantum technologies are motivated by the fact that precise tweaking of systems at a quantum level can offer several advantages over classical techniques \cite{10.2307/3559215, Schleich2016}. A particularly enticing system for the manifestation and implementation of quantum technologies are Rydberg atoms \cite{Gallagher_1988}. These are atoms in which one or more electrons have been excited to a high principal quantum number ($n$) state \cite{10.1088/978-0-7503-1635-4, Kleppner1986, Lim2013}. Because of the high $n$ state, these atoms possess some unique and exaggerated atomic properties that can be maneuvered by the selection of states and application of external electromagnetic fields \cite{doi:10.1119/1.16876}. This makes Rydberg atoms an ideal building block for quantum engineering. 

Quantum-limited field sensing and metrology are gaining immense attention in the scientific and technological community \cite{Marciniak2022, doi:10.1116/5.0007577}. \textit{``Quantum sensing"} describes the use of a quantum system, quantum mechanical phenomena, and quantum properties to perform the measurement of a physical quantity \cite{RevModPhys.89.035002}. The specialty lies in the fact that the measurement is limited only by the relevant quantum properties and not by the classical macroscopic properties of the system. Historical examples of quantum sensors include magnetometers and electrometers based on quantum interferometers and atomic vapors, or atomic clocks \cite{PhysRevLett.111.063001, Budker2007, doi:10.1063/1.2812121}. In recent years, quantum sensing has become a distinct and rapidly growing branch of research and is expected to present new opportunities -- especially with regard to high sensitivity and precision.

External oscillating fields can modulate atomic states, and the modulation is particularly stronger for Rydberg states, owing to the critical dependence of the polarizability and dipole moment on $n$ \cite{budker_kimball_demille_2008, PhysRevA.98.052503,moha08}. Rydberg Stark spectra have been investigated for rubidium with DC Stark shift using electromagnetically induced transparency (EIT). This work involves the frequency modulation of Rydberg states in thermal rubidium vapor by external electric fields oscillating at radio frequencies (RF). The modulated energy spectrum is probed with EIT \cite{10.1007/978-1-4757-9742-8_36, PhysRevLett.64.1107, krishna2005high, PhysRevA.60.2540} which is a non-destructive technique to detect the Rydberg states \cite{moha07}. We provide an experimental demonstration of the generation of sidebands in the EIT spectrum, for $^{87}\text{Rb}$, due to external oscillating fields. Studying the features of the modulated EIT spectrum provides an estimate of the strength and frequency of the external RF field. The variation of the strength of sidebands with RF voltage turns out to be a function of the Bessel functions of modulation index. Further, by fitting the experimental data with the theoretical model, we provide an estimate of the atomic polarizability associated with the Rydberg state under consideration. Authority over manipulation of Stark shifted Rydberg states allows the use of Rydberg atoms for various applications \cite{dasc12,Comp10,bara14,ande13}.

\section{Theoretical model}\label{sec2}
The shift in atomic energy levels due to the presence of an external static electric field – known as the DC Stark effect – is well known. Similarly, atomic states are also affected by a time-varying electric field, known as the AC Stark effect \cite{BAKOS1977209}.\\
\\
\textit{\textbf{Energy level shifts due to oscillating external electric field:}}
Let us consider an atom with two states $\ket{a}$ and $\ket{b}$ separated by energy $\omega_0$ and coupled by transition electric dipole moment $d$. The system is exposed to a sinusoidally varying electric field $\mathcal{E}_0\sin \omega_mt$. The relaxation rates $\Gamma$ for the states are substantially lower than all other relevant frequencies in the problem. Hence, the linewidths of the energy levels can be neglected. The amplitude of the applied field $\mathcal{E}_0$ is sufficiently small, so the effect of the electric field is treated as a weak perturbation.

To describe the energy spectrum corresponding to the system under consideration, it is helpful to consider how the energy difference between the states leads to a time-dependent quantum mechanical phase. The atomic states here are visualized as oscillators whose frequencies are modulated by the Stark effect. The energy spectrum is sketched in terms of sidebands whose relative amplitudes are given by the Bessel functions $J_n(\beta)$, where  $\beta$ is the modulation index following the expression
\begin{equation}
\label{eqn:1}
    \exp(i\beta\sin\theta)=\sum_{n=-\infty}^\infty J_n(\beta)\exp(in\theta).
\end{equation}
When the field varies with frequency $\omega_m \ll \omega_0$, the electric field at time close to $t=0$ is effectively DC. Thus, the energy-level shifts are simply given by the DC Stark shift formula:
\begin{equation}
    \Delta E=\pm\frac{d^2\mathcal{E}_0^2}{\hbar\omega_0}\sin^2\omega_mt,
\end{equation}
where the plus and minus signs refer to the upper and lower shifted states respectively. The average value of the energy shift is given by 
\begin{equation}
    \Delta E\approx\pm\frac{d^2\mathcal{E}_0^2}{2\hbar\omega_0}.
\end{equation}
The instantaneous energy of the upper state $\ket{b}$ is
\begin{equation}
\begin{split}
    E_b(t)&=\hbar\omega_0 + \Delta E(t)\\
    &=\hbar\omega_0 + \Omega\sin^2\omega_mt\\
    &=\hbar\omega_0 + \frac{\Omega}{2}-\frac{\Omega}{2}\cos2\omega_mt
\end{split}
\end{equation}
where $\Omega=d^2\mathcal{E}_0^2/\hbar\omega_0$. The phase $\phi(t)$ acquired by the state $\ket{b}$ in its time evolution is
\begin{equation*}
    \phi(t)=\int_{0}^{t} \omega(t') dt' 
\end{equation*}
Using Eqn. (\ref{eqn:1}), the time-dependent state $\ket{\psi_b(t)}=e^{-i\phi(t)}\ket{b}$ is expressed as 
\begin{equation}
\begin{split}
    \label{eqn:fmwf}
    \ket{\psi_b(t)}&=\exp{-i(\omega_0+\Omega/2)t}\ket{b}\times\\
    &\sum_{n=-\infty}^\infty J_n\left(\frac{\Omega}{4\hbar\omega_m}\right)\exp(i2n\omega_mt)
\end{split}
\end{equation}
Thus, it can be observed that the energy spectrum is given by a set of sidebands centered around $\omega_0+\Omega/2$ and spaced by frequency intervals of $2\omega_m$. The statistical weight of each energy sideband, which is a measure of the strength (or intensity) of a sideband, is given by the square of the Bessel functions $J_n(\beta)$ where the modulation index $\beta=\Omega/4\hbar\omega_m$.

In case of frequency modulation of light waves, a pair of sidebands centered around the carrier frequency is obtained for low modulation strength \cite{Kotov1997, 8943124}. Similarly, frequency modulation in atomic states also leads to the generation of sidebands at frequencies $\pm2\omega_m$ from the unmodulated signal frequency, subject to the condition $\beta\ll1$. The relative intensity of the sideband (w.r.t. the central energy peak) is given by $J_1(\beta)^2$. For higher modulation strengths, it is possible to obtain the higher-order sidebands with relative intensities $J_n(\beta)^2$, where $n=2,3,...$ denotes the order of the sidebands.

\section{Methods}\label{sec3}
\begin{figure}
    \center
    \includegraphics[width=1\linewidth]{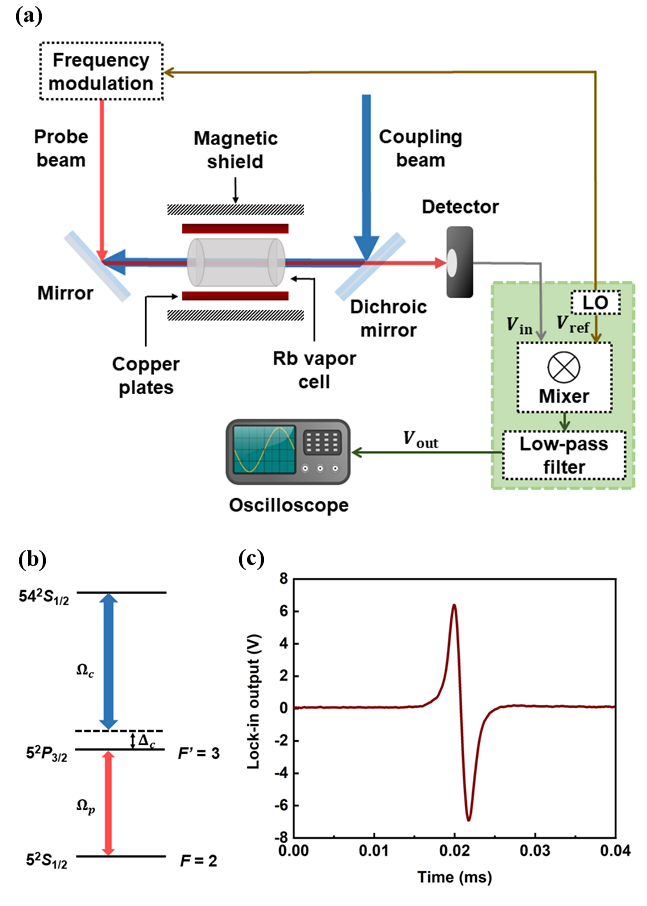}
    \caption{(a) Schematic of the experimental setup. (b) Energy level scheme corresponding to the formation of EIT in $^{87}\text{Rb}$. (c) Probe transmission intensity in absence of any external field.}
    \label{fig:Fig1}
\end{figure}

The schematic of the experimental setup is shown in Fig. \ref{fig:Fig1} (a). 
A probe beam at 780 nm is derived from a diode laser which couples the ground state $5S_{1/2}$ to the excited state $5P_{3/2}$. The probe beam is split into two parts. One part of the beam is used to stabilize its frequency at the transition $^{87}\text{Rb}$ $5S_{1/2}$ $F=2\longrightarrow 5P_{3/2}$ $F'=3$. The other part of the probe beam is used for achieving EIT. A coupling beam at 480 nm is derived from a frequency doubling cavity laser to couple the excited state $5P_{3/2}$ to the Rydberg state $54S_{1/2}$. $\Delta_c$ denotes the detuning of the coupling beam frequency from resonance. The probe and the coupling beams are made to counter-propagate through a rubidium vapor cell. The probe transmission signal is observed using a photodetector with the coupling beam scanned across resonance to observe EIT. The energy level scheme showing the Rydberg excitation corresponding to the formation of dark states in $^{87}\text{Rb}$ is shown in Fig. \ref{fig:Fig1} (b).
The power of the probe (coupling) laser beam used for the experiment is $\approx$ 1.5 $\mu$W ($\approx240$ mW). The $1/e$ radii of the probe and coupling beams are 46.04 $\mu$m  and 44.56 $\mu$m respectively.

To achieve a good signal-to-noise ratio (SNR), a lock-in detection is performed. The probe beam is frequency modulated through the current modulation unit of the diode laser which is operated using the local oscillator of a lock-in amplifier at a frequency of 100 kHz. The frequency modulated signal observed at the photo-detector is multiplied with an internal reference signal of the lock-in amplifier. The internal reference is produced from the same local oscillator with a frequency equal to the modulation frequency. The resultant signal passes through an integrated low-pass filter. The final output signal has dispersion signal-like behavior with an enhanced SNR as shown in Fig. \ref{fig:Fig1} (c). To observe the effect of RF electric fields, a pair of rectangular copper plates having a cross-sectional area of $13.5\times3$ cm$^2$ is placed parallel to the longitudinal side of the absorption cell. The separation between the plates is 5.5 cm.

\section{Results}\label{sec4}
\begin{figure}
    \center
    \includegraphics[width=\linewidth]{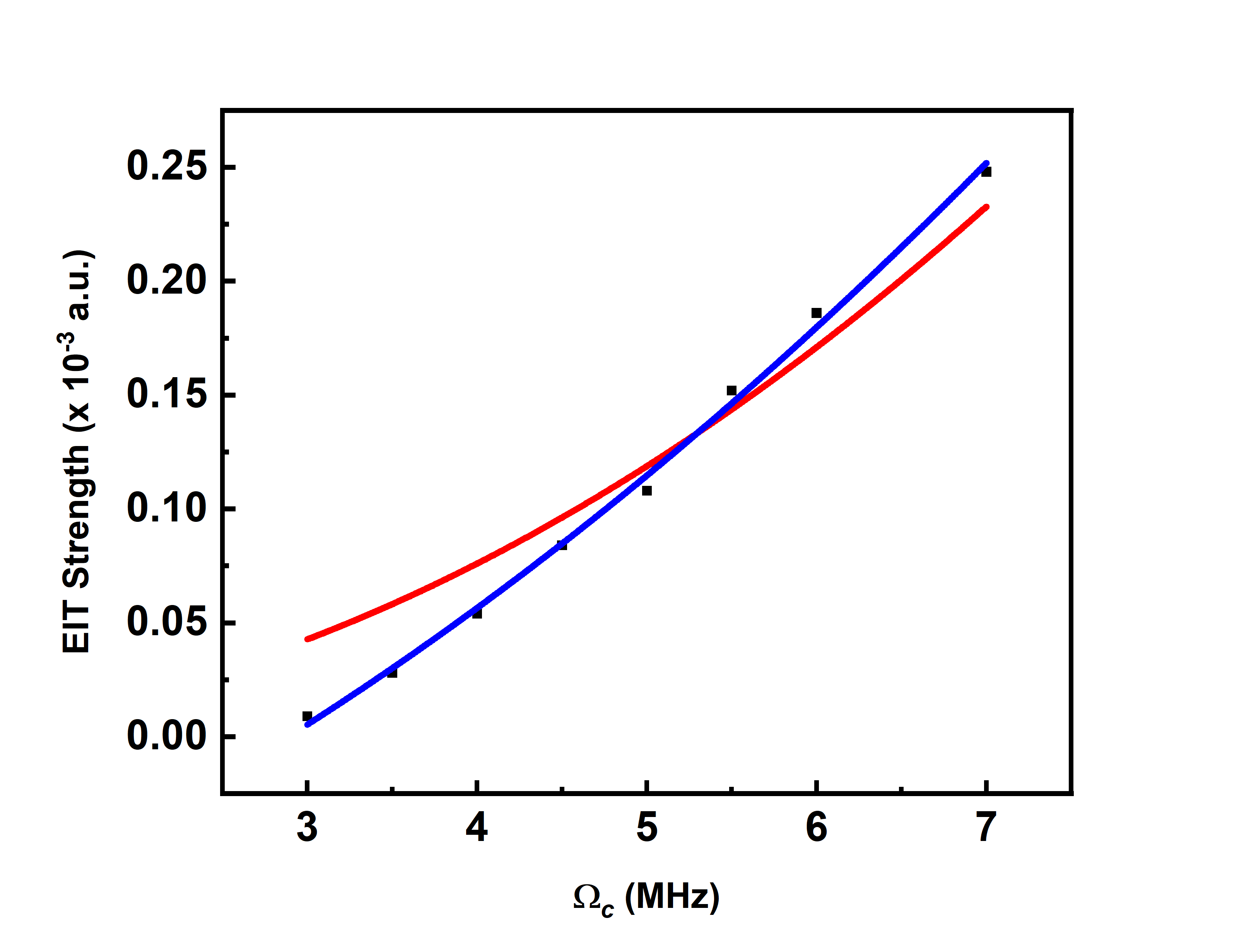}
    \caption{Simulated EIT strength as a function of $\Omega_c$. Red curve: The fitting function used is $y=a_1\cdot x^2$. The first term of Eqn. \ref{eqn:7} gives the Doppler-broadened absorption profile, and is not considered in the fitting function used to calculate the EIT strength. For low values of $\Omega_c$, the third and higher terms can be neglected. Blue curve: The fitting function used is $y=b_1\cdot x+b_2\cdot x^2$.}
    \label{fig:Fig2}
\end{figure}
In this work, EIT is implemented as a technique to probe the Rydberg energy states arising due to the modulation by external RF fields. The strength of the EIT peak, measured in terms of the imaginary part of susceptibility $\chi^{\text{Im}}$, depends on the coupling Rabi frequency $\Omega_c$ according to the relation \cite{PhysRevA.46.R29}
\begin{equation}
\begin{split}
    \chi^{\text{Im}}&\propto\frac{1}{\Omega_c^2+\Gamma_2\Gamma_3}\\
    &=\frac{1}{\Gamma_2\Gamma_3}\left(1+\frac{\Omega_c^2}{\Gamma_2\Gamma_3}\right)^{-1}
\end{split}
\end{equation}
where $\Gamma_2$ and $\Gamma_3$ denote the decay rates of states $\ket{2}$ and $\ket{3}$ respectively. For suitable values of $\Omega_c$, $\Gamma_2$ and $\Gamma_3$, the binomial expansion of the expression in parentheses in the preceding equation gives:
\begin{equation}
\label{eqn:7}
    \chi^{\text{Im}}\propto 1-\frac{\Omega_c^2}{\Gamma_2\Gamma_3}+\frac{\Omega_c^4}{(\Gamma_2\Gamma_3)^2}-\frac{\Omega_c^6}{(\Gamma_2\Gamma_3)^3}+...
\end{equation}
Corresponding to the parameters used in this study, the variation of EIT strength as a function of $\Omega_c$ is shown in Fig. \ref{fig:Fig2}. It is observed that for low values of $\Omega_c$, the fitting function (red curve) inspired from Eqn. (\ref{eqn:7}) deviates significantly from the simulated values of EIT strength. This is attributed to enhanced absorption occurring at low values of $\Omega_c$ \cite{Bae:10}. To account for this enhancement of absorption, we fit the EIT strengths corresponding to the low-lying values of $\Omega_c$ to a function that contains a term linear in $\Omega_c$ in addition to a quadratic term. The modified fitting is shown in blue in Fig. \ref{fig:Fig2}.

\begin{figure}
    \center
    \includegraphics[width=\linewidth]{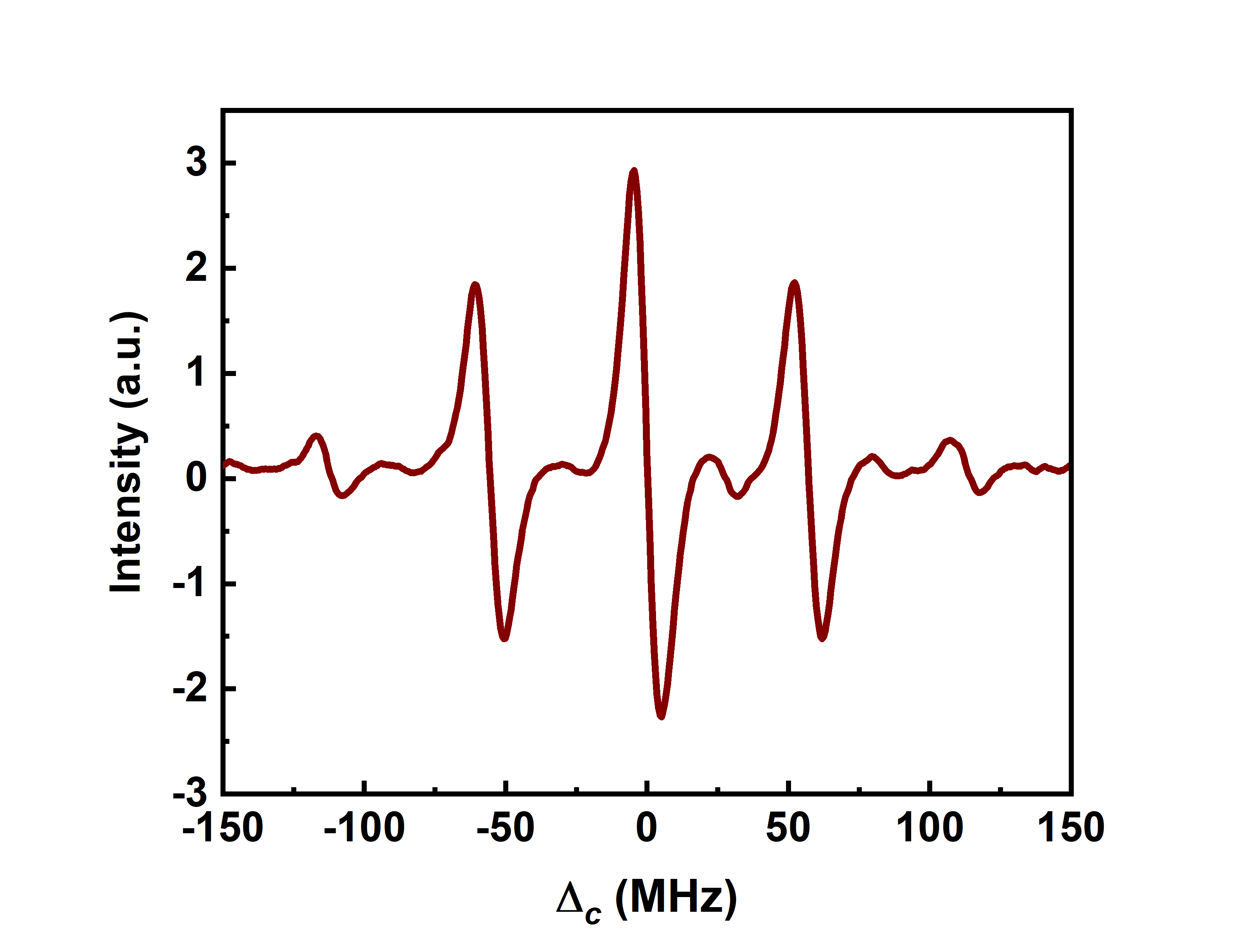}
    \caption{Probe transmission as a function of $\Delta_c$ in presence of an RF field.}
    \label{fig:eit28MHz}
\end{figure}
In the presence of external RF fields, the coupling Rabi frequencies are scaled by appropriate Bessel functions in accordance with the analysis in Section \ref{sec2} and the modified fitting in Fig. \ref{fig:Fig2}. The EIT strength $S_n$ corresponding to the $n^{\text{th}}$ sideband, for the coupling Rabi frequencies considered in this study, is thus approximated as
\begin{equation}
    S_n\approx k_1\cdot J_n(\beta)+k_2\cdot J_n(\beta)^2
\end{equation}
where $\beta=\alpha V_0^2/4\hbar\omega_m l^2$. $\alpha=d^2/\hbar\omega_0$ is the atomic polarizability, $V_0$ is the peak RF voltage and $l$ is the distance between the capacitor plates.

To demonstrate the effect of an RF electric field on the EIT spectrum, electric fields of several amplitudes and frequencies in the radio-wave band are applied to the atomic system. The EIT spectrum in the presence of an electric field of amplitude 7.2 V and frequency 28 MHz is shown in Fig. \ref{fig:eit28MHz}. In addition to the main EIT peak, which has been calibrated to zero frequency, a clear presence of two pairs of sidebands at frequencies close to -55.8 MHz, +56.8 MHz and -112.6 MHz, +112.6 MHz are observed. It is verified that these sidebands occur at the second harmonics ($\pm2\nu$) and the fourth harmonics ($\pm4\nu$) of the applied field frequency ($\nu$) respectively. 

\begin{figure}
    \center
    \includegraphics[width=0.95\linewidth]{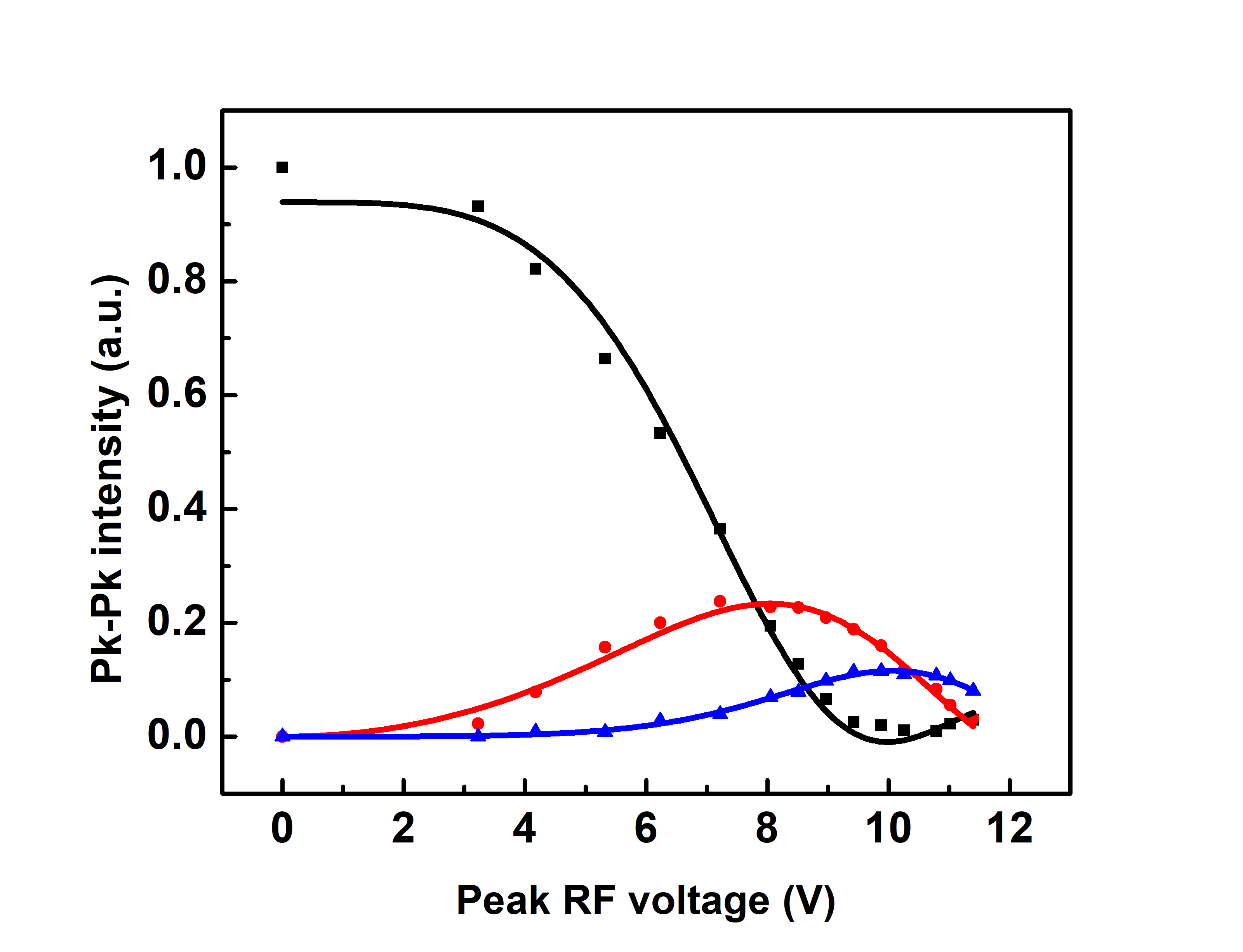}
    \caption{Peak-to-peak intensity of central EIT peak (black squares), first sideband (red circles) and second sideband (blue triangles) obtained from experiment. Their best fit to the Bessel function is represented by the corresponding solid lines.}
    \label{fig:bessel_fit}
\end{figure}
For further analysis, the frequency of the applied field is kept constant at 28 MHz and the amplitude (peak RF voltage) is varied in the range of 0-11.4 V. From each of these spectra, the peak-to-peak intensities of the sidebands are extracted and plotted as a function of varying RF voltage. The resultant plots are shown in Fig. \ref{fig:bessel_fit} where the scattered plots correspond to the experimental data. These scattered plots are then fitted with the function $S_n$, for $n=0, 1, 2$ with $k_1$, $k_2$ and $\alpha$ as the fitting parameters. The fitted curves are plotted in solid lines in the graph shown in Fig. \ref{fig:bessel_fit}. The value of atomic polarizability obtained from the fitting is $\alpha=(h)\cdot1.01\times10^8$ $\text{Hz/(V/cm)}^2$ with a root-mean-square error of 1.64 \%, which is close to the value of polarizability of $54S_{1/2}$ state reported by Lai \textit{et al.} \cite{PhysRevA.98.052503}.

\section{Conclusion}\label{sec5}
We demonstrate the frequency modulation of the Rydberg states in thermal atomic vapor due to the presence of an oscillating RF electric field. The modulation results in the generation of sidebands around the EIT signal. The relative frequency of the generated sidebands gives an estimate of the frequency of the RF field. The peak-to-peak amplitudes of the EIT signal and the sidebands fit quantitatively to a combination of Bessel functions of the modulation index which is used to determine the atomic polarizability of Rydberg states. This study proves to be an efficient technique to detect RF fields and quantify the effect of RF fields on Rydberg states in thermal atomic vapor which can be used for precise quantum sensing applications.

\section{Acknowledgement}\label{sec7}
The author gratefully acknowledges the financial support from the Department of Atomic Energy, Government of India under the Project Identification No. XII-R$\&$D-5.02-0200 (National Institute of Science Education and Research Bhubaneswar) and funding in the form of a fellowship from CHANAKYA-PG Fellowship 2022. The author is also grateful to Dr. Ashok K. Mohapatra, Associate Professor, School of Physical Sciences, National Institute of Science Education and Research Bhubaneswar for his able guidance at several stages of the work pertaining to this paper, and Dr. Tanim Firdoshi, Dr. Suman Mondal, Mr. Sujit Garain for their assistance in building the experimental setup and inspiring discussions on various aspects of the work. The author would like to extend gratitude to Prof. Nick Vamivakas, Professor of Quantum Optics and Quantum Physics, University of Rochester for his valuable inputs.

\bibliography{ref}
\bibliographystyle{ieeetr}

\end{document}